\documentclass[12pt,preprint]{aastex}

\newcommand{\kms}   {~km~s$^{-1}$}
\newcommand{\mjy}   {~mJy~beam$^{-1}$}
\newcommand{\jy}    {~Jy~beam$^{-1}$}

\newcommand{\cmt}   {~cm$^{-3}$}
\newcommand{\vlsr}  {$v_{\rm LSR}$}

\newcommand{\mo}    {$M_{\sun}$}

\newcommand{\nh}    {NH$_3$}

\newcommand{\form}  {H$_2$CO}

\newcommand{\J}[2]  {\mbox{#1--#2}}
\newcommand{\JK}[4] {\mbox{#1$_{#2}$--#3$_{#4}$}}

\slugcomment{Submitted to the \apj}

\shorttitle{Resolving a double core in L723}
\shortauthors{Girart et al.}

\begin{document}


\title{The L723 low mass star forming protostellar system: resolving a double
core}


\author{J. M. Girart}
\affil{Institut de Ci\`encies de l'Espai, (CSIC-IEEC), 
Campus UAB, Facultat de Ci\`encies, Torre C5 - parell 2, 
08193 Bellaterra, Catalunya, Spain}
\email{girart@ieec.cat}

\author{R. Rao}
\affil{Submillimeter Array, 
Academia Sinica Institute of Astronomy and Astrophysics,
645 N. Aohoku Pl, HI 9672}
\email{rrao@sma.hawaii.edu}

\and

\author{R. Estalella}
\affil{Departament d'Astronomia i Meteorologia, Universitat de Barcelona, 
Mart\'i i Franqu\`es 1, 08028 Barcelona, Catalunya, Spain}


\begin{abstract}
We present 1.35 mm Submillimeter Array (SMA) observations around the low--mass
Class 0 source IRAS 19156+1906, at the  the center of the LDN~723 (L723) dark
cloud.   We detected emission from dust as well as emission from \form\
\JK{3}{0,3}{2}{0,2}, DCN \J{3}{2} and CN \J{2}{1} lines, which arise from two
cores, SMA~1 and SMA~2, separated by $2\farcs9$ (880 AU in projected
distance).  SMA~2 is associated with the previously detected  source VLA~2.
Weak SiO \J{5}{4}\ emission is detected, possibly tracing a region of
interaction between the dense envelope and the outflow.  We modeled the dust
and the \form\ emission from the two cores. The results from the modeling show
that the cores have similar physical properties (density and temperature
distribution) but  that SMA~2 has a larger p--\form\ abundance (by a factor
3--10) than SMA~1. The p--\form\ abundances found  are compatible with the
value of the outer part of the circumstellar envelopes associated with Class 0
sources. SMA~2 is harboring an active multiple low--mass protostellar system
and powering at least one molecular outflow \citep{Carrasco08}. In contrast,
there are no known signs of outflow activity towards SMA~1.  This suggests that
SMA~2 is more evolved than SMA~1. The kinematics of the two sources show
marginal evidence of infall and rotation motions. The mass detected by the SMA
observation, which trace scales of $\la$1000~AU, is only a small fraction of
the mass contained in the large scale  molecular envelope, which suggests that
L723 is still in a very early phase of star formation. Despite the apparent
quiescent nature of the L723, fragmentation is occurring at the center of the
cloud at different scales. Thus, at $\simeq1000$~AU the cloud has fragmented in
two cores, SMA 1 and SMA 2.   At the same time, at least one of these cores,
SMA~2, has undergone additional fragmentation at scales of $\simeq150~$AU,
forming a multiple stellar system.  
\end{abstract}



\section{Introduction}

L723 is an isolated molecular cloud located at a distance of $300 \pm 150$~pc
\citep{Goldsmith84}  and with a systemic velocity of  $V_{\rm LSR} \simeq
10.9$\kms\ \citep{Girart97}.  It harbors a low--mass, Class 0, young stellar
object, first identified  by the IRAS satellite, IRAS 19156+1906
\citep{Goldsmith84}, with a bolometric luminosity of 3.4~L$_{\odot}$
\citep{Dartois05}. The  properties of the protostar  in the IR to mm wavelength
range have been object of several studies \citep{Davidson87, Andre93,
Pezzuto02, Dartois05}. IRAS 19156+1906 is associated with a CO outflow with a
quadrupolar morphology \citep{Goldsmith84, Moriarty89, Avery90, Hayashi91,
Lee02}. This outflow consists of a pair of bipolar lobes aligned along the
East--West direction (position angle $PA\simeq100\arcdeg$) and another pair of
bipolar lobes aligned roughly in the North-South direction
($PA\simeq32\arcdeg$), with IRAS 19156+1906 located at their common center 
\citep{Avery90, Lee02}.  The East--West (EW) pair of lobes, and in particular
the blueshifted, eastern lobe, is associated with several Herbig-Haro (HH)
objects \citep{Palacios99, Lopez07}. Very Large Array (VLA) observations at
3.6~cm reveal two sources, VLA~1 and VLA~2,  toward the center of outflow
\citep{Anglada91}, although VLA~1 is likely a background source
\citep{Anglada96,Girart97}.  VLA~2 shows the  characteristics of a thermal
radio-jet  (jet--like morphology and partially optically thick free-free
emission),  and was first identified as the powering source of the  EW pair of
molecular lobes \citep{Anglada96}. Recent, very sensitive, sub-arcsecond
angular resolution  ($0\farcs2$--$0\farcs7$) VLA observations at 3.6 cm and 7
mm resolve VLA 2 into several components. The two brightest sources at 3.6~cm,
VLA~2A and 2B, are separated by $0\farcs3$ (90 AU in projection) and are
possibly tracing embedded protostars \citep{Anglada04,Carrasco08}. VLA~2A is
associated with extended emission along a $PA \simeq 115\arcdeg$.  Two
additional sources, believed to also trace embedded protostars are identified
at 7~mm: VLA~2C, located at the position of the water maser emission
\citep{Girart97,Furuya03}, and VLA~2D, located $\sim 3''$ southeast of VLA~2A.
\citet{Carrasco08} suggest that the CO high velocity emission and the HH
objects are tracing three  different outflows. One of them is the NS outflow
possibly powered by VLA~2B.  The other two are what previously was considered
as the EW pair of lobes: one of  the outflows has a $PA\simeq 115\arcdeg$ and
is powered by VLA~2A, whereas the other one seems to be  a ``fossil'' outflow
with a $PA\simeq 90\arcdeg$. 

High angular resolution (3\farcs5) VLA observations of L723 carried out by
\citet{Girart97}, show that the \nh\ emission arises from a V-shaped structure
that traces the dense molecular envelope around the embedded protostars. This
structure is elongated roughly east-west and is 0.15~pc long, with a mass of
$7~M_{\odot}$. The ammonia maps show evidence of heating and line broadening
toward VLA 2. A second spot of heating is observed $10\arcsec$ west of VLA 2.
This western hot spot (WHS here after) is interpreted by \citet{Girart97} as a
very young protostar with and age shorter than $2\times10^{4}$~yr. The \nh\
structure is only partially traced by CS and N$_2$H$^+$
\citep{Hirano98,Chen07}. The dense envelope of L723 was also observed at
sub-millimeter wavelengths ($850~\mu$m and $450~\mu$m) by \cite{Shirley02} and
\citet{Estalella03} using the continuum SCUBA camera of the JCMT. A strong
millimeter source was found, peaking at the position of VLA 2, with a large
extension that matches the high-density gas traced by \nh. 

In this paper we present a study of the molecular and dust emission at scales
of $\sim 1000$~AU around the low--mass star forming region of L723. In \S~2 we
describe the observational procedure with the SMA. In \S~3 we describe the main
results obtained with the SMA data. In \S~4 we present an analysis of the dust
and \form\ emission. In \S~5 we discuss the possible evolutionary scenarios of
the sources in the region.

\section{SMA Observations}\label{Sobs}

The observations were carried out with the 8--antenna SMA\footnote{The
Submillimeter Array is a joint project between the Smithsonian Astrophysical 
Observatory and the Academia Sinica Institute of Astronomy and Astrophysics, 
and is funded by the Smithsonian Institution and the Academia Sinica.} array
at Mauna Kea at 1.3 millimeters in 2004 August (compact
configuration).  The phase center was  set at  
$\alpha(J2000)=19^\mathrm{h}17^\mathrm{m}53\fs400$  and
$\delta(J2000)=19\degr12'19\farcs40$.  The phase calibrator used was QSO
1925$+$211, and QSO 1749$+$096 was used as a reference calibrator. 
Absolute flux and bandpass calibration was done by observing Uranus and Jupiter,
respectively. 
The SMA correlator has a bandwidth of 2~GHz with 24
partially overlapping windows (chunks) of 104~MHz. All the chunks had 128
channels, except the second chunk which was configured to have 512 channels.  
The correlator was setup to observe the \form\ \JK{3}{0,3}{2}{0,2} line in
the  second chunk of the lower side band.  Maps were made with the $(u,v)$
data weighted by the associated system  temperatures and using a robust
weighting of 0.0 (continuum data), 0.5 (\form)  and 1 (other line data).
Table~\ref{tobserva} lists the basic information about the resulting maps, 
including the frequency of the lines or continuum, the channel resolution 
(for line observations), the resulting synthesized beam and the $rms$ noise of
the maps.

\section{Results}

\subsection{Dust continuum\label{ResultDust}} 

\begin{figure}
\epsscale{0.8}
\plotone{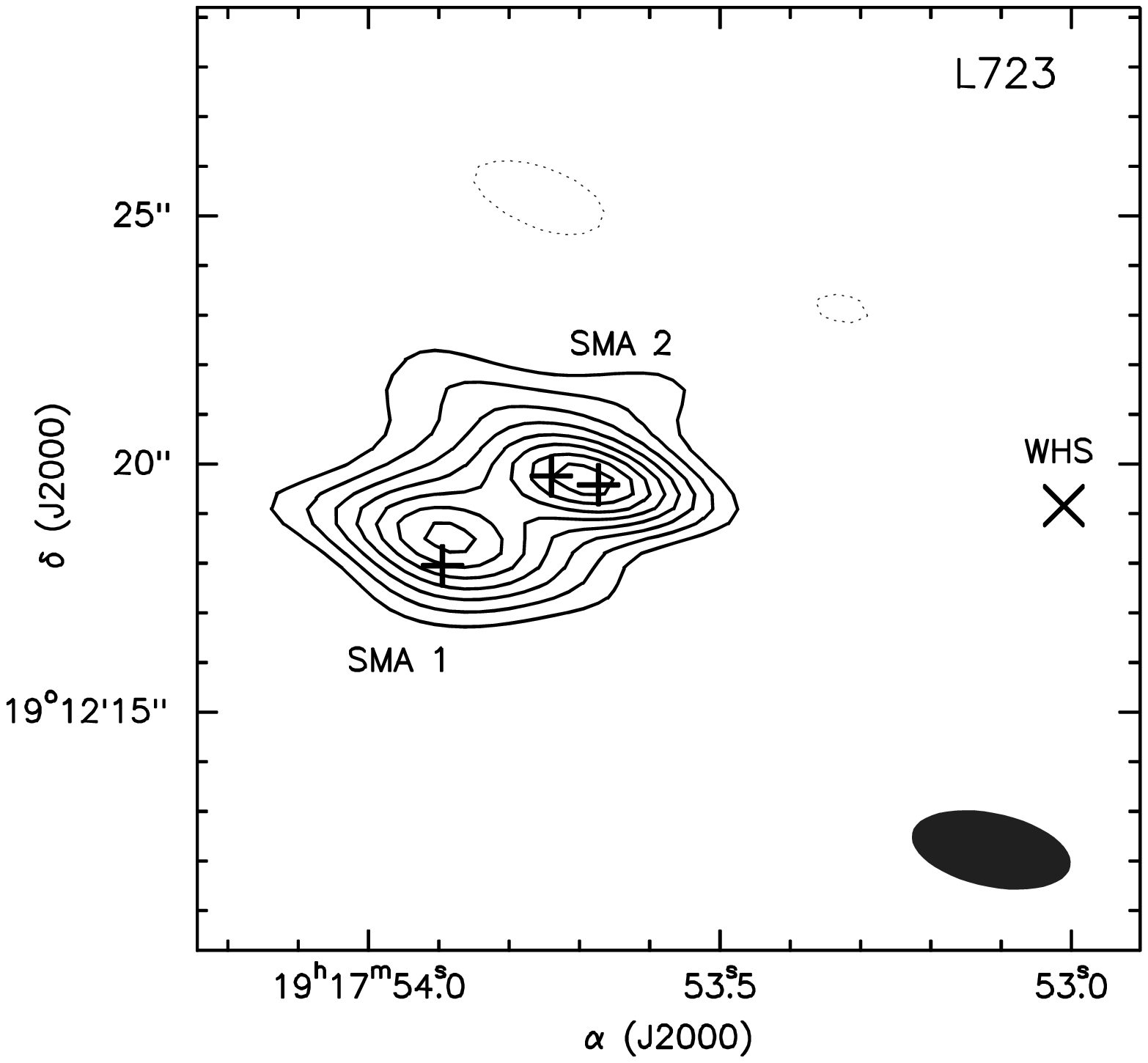}
\caption{
SMA map of the continuum emission at 1.35~mm (222.31 GHz) toward L723 VLA~2. 
Contour levels are $-3$, 3, 5, 7, 9 , 11, 13, 15 and 17 times 1.8~\mjy, the
rms noise  of the map. The synthesized  beam is shown in the lower right
corner.  The crosses show the position of  VLA~2A (the western cross) and
VLA~2C,  which are associated with SMA 2, and VLA~2D, associated with SMA 1.
These sources were detected by the VLA at 3.6~and 7~mm from sub--arcsecond
maps by \citet{Carrasco08}. The titled cross shows the position of the WHS
\citep{Girart97}.
\label{fdust}}
\end{figure}

The 1.35~mm continuum map shows that the emission is clearly resolved into two 
components, L723 SMA 1 and SMA 2, with similar intensities (see
Figure~\ref{fdust})  and  separated by $2\farcs9$ (880 AU in projected distance)
with a position angle of $106\arcdeg$. These two components were previously
detected by \citet{Launhardt04} at 3~mm. The strongest component, SMA~2,  is
associated with VLA 2A, 2B and 2C, 3.6~cm and 7~mm continuum sources  that are
probably young stellar objects \citep{Carrasco08}. The other component,  SMA~1,
is only detected at 3~mm \citep{Launhardt04} and marginally at 7~mm (source VLA
2D: Carrasco et al. 2008) and does not have known counterparts at other
wavelengths. Table~\ref{tdust} gives the values  of position, intensity, flux
densities and   the deconvolved sizes for both mm sources obtained by fitting a
Gaussian to each component.  The two sources are partially resolved at scales of
about  $1\farcs0$--$2\farcs0$  (300--600~AU). 

Naturally weighted maps were obtained in order to better estimate the total flux
of the double core  and to try to detect the WHS.  The total flux density 
detected by our observations for SMA~2 and SMA~1 is $128\pm6$~mJy, which is
22~mJy  higher ($\sim$20\%) than that from the map shown in Figure~\ref{fdust}, 
obtained with a robust weight of 0. This excess  is possibly due to the better
sensitivity of the natural weighted map to weak and extended emission.  No
emission is detected with the SMA at the position  of WHS, with a 3-$\sigma$
upper limit of 4.2~\mjy. Gaussian tapers  were applied to the visibilities to
increase the beam size and see if the WHS could be  detected, which yielded
negative results. 

The mass of the circumstellar material traced by the dust emission at mm
wavelength, where it is optically thin is:
\begin{equation}\label{equ}
M = \frac{gS_{\nu}D^2}{\kappa_{\nu}B_{\nu}(T_{\rm dust})}
\end{equation}
with $g$ the gas--to--dust ratio, $D$ the distance, $S_{\nu}$ the flux density,
$\kappa_{\nu}$ the dust mass opacity coefficient and $B_{\nu}$ the Planck 
function for a blackbody of dust temperature $T_{\rm dust}$. The major source  
of uncertainty in the mass derivation is the poor knowledge  of the distance
($300\pm150$~pc).  To derive the mass  we adopted $g=100$ \citep{Draine04},  a
dust mass opacity of  $\kappa_{\rm 250GHz}=0.9$~cm$^2$~g$^{-1}$ 
\citep{Ossenkopf94}. To estimate the temperature of the dust, we used the VLA 
NH$_{3}$ (1,1) and (2,2) maps from \citet{Girart97}, which have a slightly 
higher angular resolution ($3\farcs7\times3\farcs3$) than the SMA maps.  
Following the standard analysis procedure for these two ammonia lines, we
obtain that at the SMA~1, SMA~2 and WHS peak positions the ammonia rotation
temperature  is  $\simeq 20$~K, which converted to the gas  kinetic temperature
of the gas  is $\sim 25$~K \citep{Danby88}.  Table~\ref{tmass} gives the mass
derived for  the different components using this value for the dust
temperature.  The  total mass of the emission detected by the SMA is $\sim
0.24$~\mo, which is only  a small fraction, $\sim 5$\%, of the dense  envelope
detected at larger scales \citep{Girart97,Shirley02}.  WHS has a very  small
mass, $\la 0.01$~\mo.

\subsection{H$_2$CO} \label{srform}

\begin{figure}
\epsscale{1.0}
\plotone{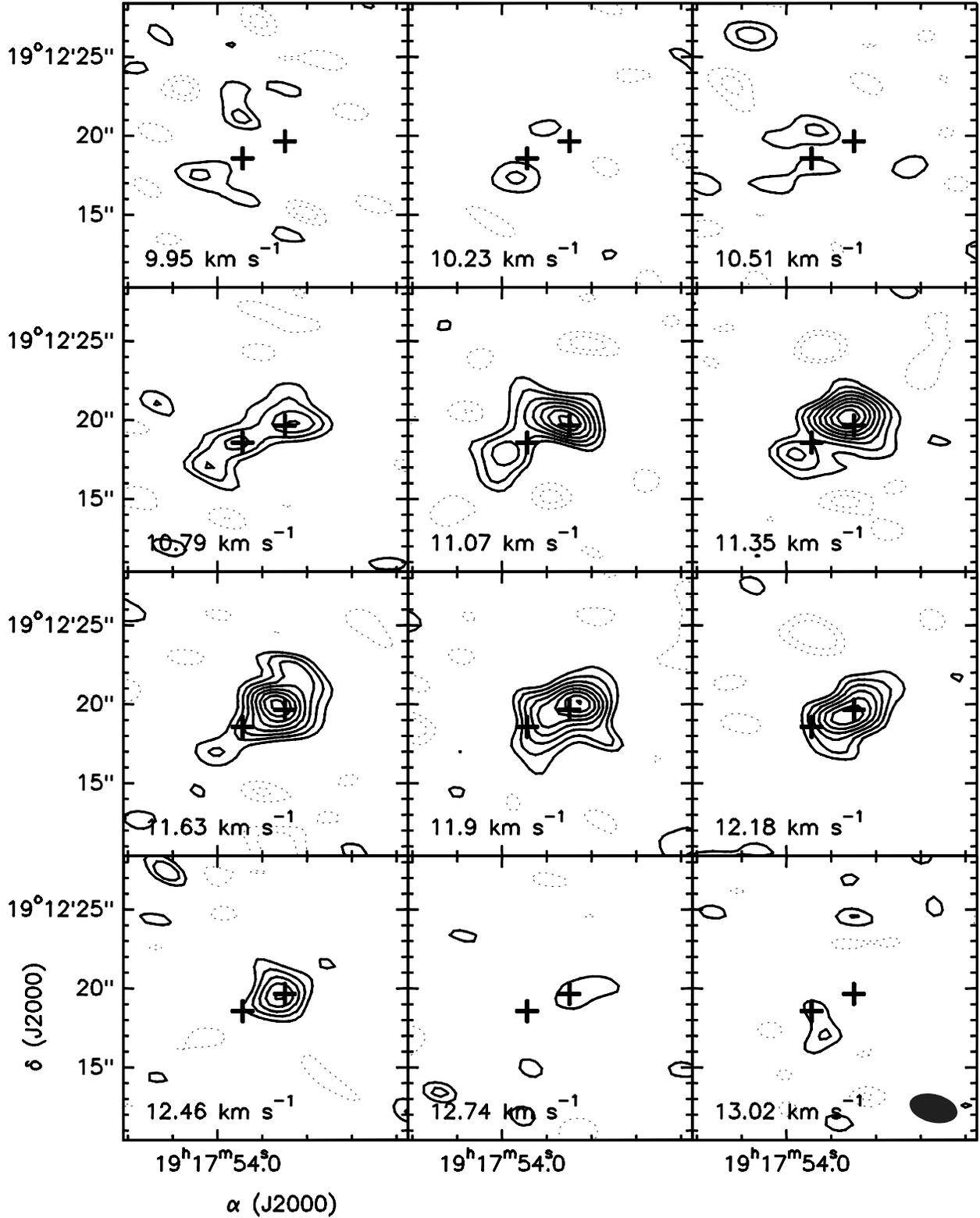}
\caption{
SMA map of the \form\ \JK{3}{0,3}{2}{0,2} emission toward L723 VLA 2.  Contour
levels are $-2$, 3, 4, 5, 6, 7 and 8 times 0.18~\jy, the rms noise of the
map. The synthesized  beam is shown in the lower right corner of the lower
right panel. The crosses show the position of the dust sources SMA~1
and SMA~2. The \vlsr\ velocity of the channels is shown in the lower left corner
of the panels. 
\label{fform}}
\end{figure}

\begin{figure}
\epsscale{0.8}
\plotone{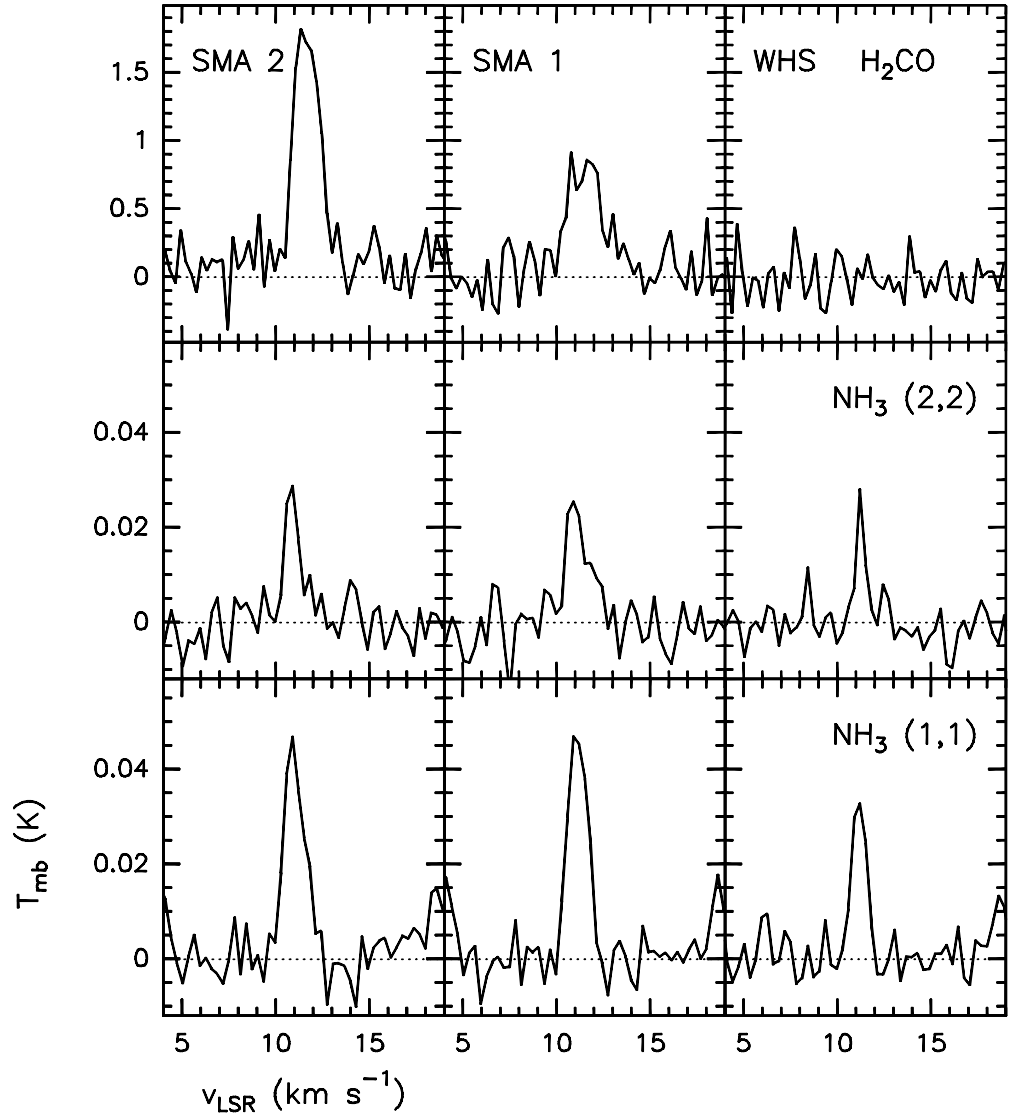}
\caption{
Spectra of the \form\ \JK{3}{0,3}{2}{0,2}, \nh\ (2,2) and (1,1) toward the
position of SMA~2 (left panels), SMA~1 (central panels) and WHS (right panels).
The spectra were taken from channel maps at similar angular resolution, $\sim
3''$. The \nh\ spectra are taken from
\citet{Girart97}.
\label{fespectra}}
\end{figure}

\begin{figure}
\epsscale{0.7}
\plotone{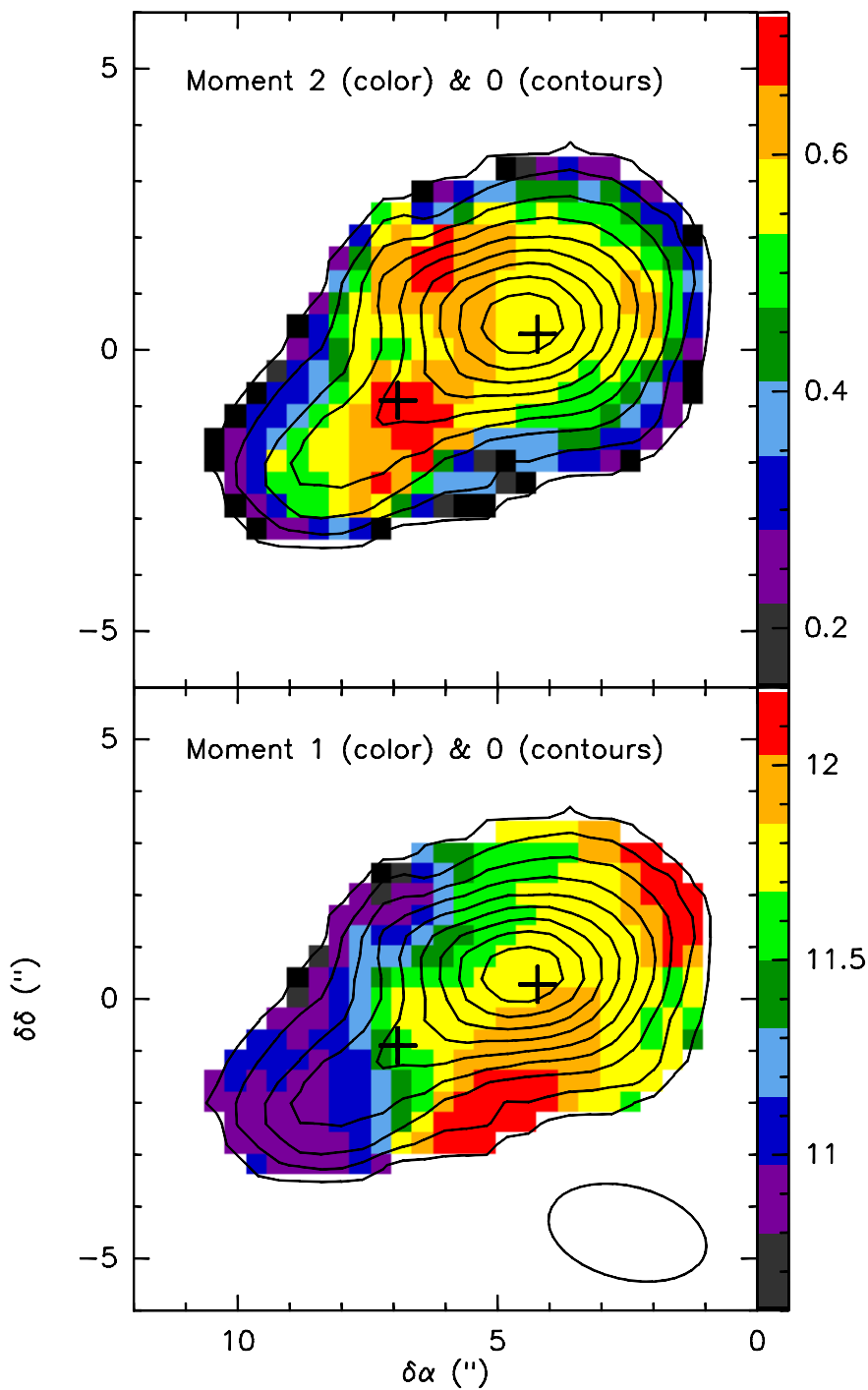}
\caption{
Color image of the first order (bottom panel) and second order (top panel)
moments of the \form\ emission towards L723 VLA~2 superposed with the zero
order  moment (integrated emission) contour map. The color image level is shown
in  the right side of the panels. The first contour is 0.1~\jy\ \kms, and the
contour  levels are 0.3 ~\jy\ \kms. The crosses show the position of the dust
sources SMA~1 and SMA~2. The synthesized beam of the maps is shown in the
bottom right corner of the bottom panel.
\label{fmom}}
\end{figure}

Figure~\ref{fform} shows the channel maps for the \form\ \JK{3}{0,3}{2}{0,2}\
line. The \form\ emission is concentrated around the two mm continuum sources,
SMA~1  and SMA~2, although in most of the channels SMA~2 has brighter \form\
emission  than SMA~1. Toward SMA~2 the \form\ emission peaks at or very close
to the dust  emission peak, whereas around SMA~1 the peak emission is, in some
channels, displaced to the SE (only at $v_{\rm LSR}=10.79$\kms\ the
\form\ peaks at the same position as the dust).  The kinematics of the two
sources is difficult to disentangle because their separation  is similar to
the angular resolution. The first order moment of the emission shows velocity
gradients along the two cores with velocity shifts up to $\sim 1$\kms. 
Although there is not a clear simple trend, the emission east of SMA~1 is
clearly  blueshifted with respect to the rest of the region. The velocity
dispersion of the  \form\ emission (second-order moment map: see
Fig.~\ref{fmom}) changes  little around the two cores, with values within the
range of 0.5  to 0.7\kms, which corresponds, for a Gaussian line, to a full
width at half maximum of 1.2 to 1.7\kms. This velocity dispersion range is
larger than the one measured from VLA NH$_3$ observations, $\le 0.42$\kms\
\citep{Girart97}.  Figure~\ref{fespectra} shows that the spectra of \form\
has larger line widths that those of \nh\ (note  that the (1,1) and (2,2)
transitions have hyperfine components, which broadens the line, specially for
the (1,1)).

\begin{figure}
\epsscale{0.55}
\plotone{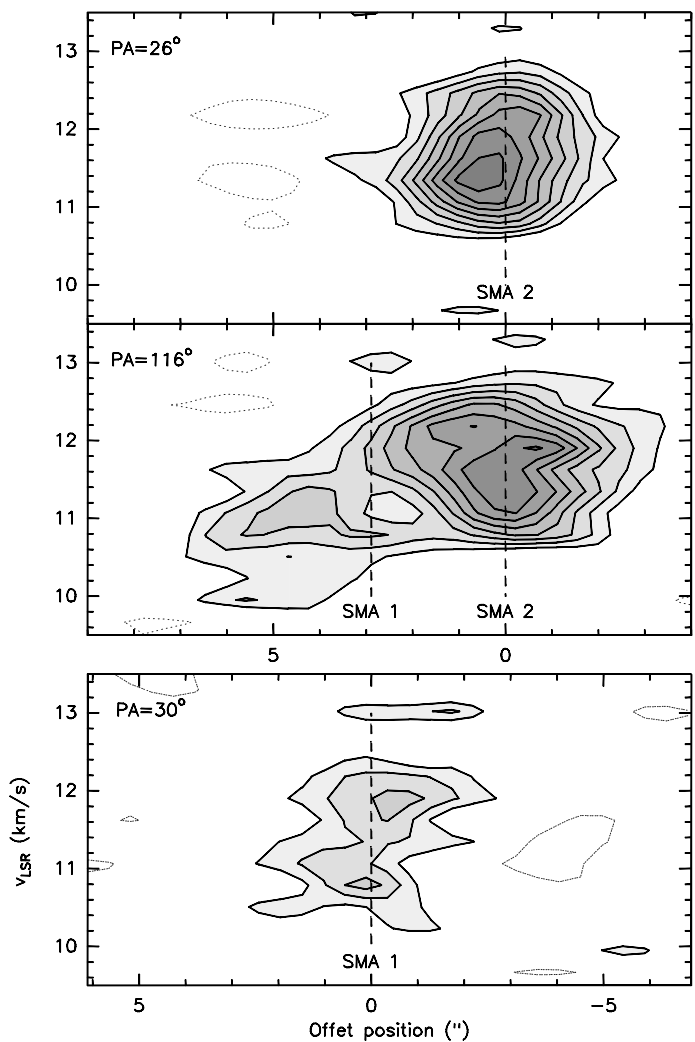}
\caption{
Position-velocity (PV) plots of the \form\ \JK{3}{0,3}{2}{0,2} emission. The
{\it top} and {\it central} panels  show the PV plot perpendicular and along 
the VLA~2 thermal radio jet direction, $PA=26\arcdeg$ and $116\arcdeg$,
respectively, centered on SMA~2 (and VLA~2). The {\it bottom} panel shows the 
PV plot centered on SMA~1 with a position angle, $PA=30\arcdeg$ nearly
perpendicular to the VLA~2 radio jet and similar to the north-south pair of
lobes of the quadrupolar molecular outflow \citep{Lee02}.
\label{fpv}}
\end{figure}

Since VLA~2  is a thermal radio jet with a well defined orientation
\citep{Anglada96}, position-velocity (PV) plots along ($PA=116\arcdeg$) and
perpendicular  ($PA=26\arcdeg$) to the radio jet orientation were made crossing
VLA~2, which lies at the peak of SMA~2  (Figure~\ref{fpv}). The PV plot along
the radio jet direction passes very close  to SMA~1 (see Fig.~\ref{fpv}).  In
the direction perpendicular to the jet,  the \form\ shows a velocity gradient
($\sim 0.8$\kms\ within $1''$ around  VLA~2) that could be indicative of
rotation. In the direction along the radio jet  there is overlap of the
emission between SMA~2 and SMA~1. Nevertheless, at  the position of SMA~1 the
\form\ emission shows a relative minimum. In Figure~\ref{fpv} we also show the
PV plot centered on SMA~1 with a position angle  of $PA=30\arcdeg$, nearly
perpendicular to the PV along the VLA~2 radio jet direction. This PV plot shows
two weak peaks at different velocities, \vlsr$\simeq 10.8$  and 11.9\kms, that
are slightly shifted in position ($\simeq 0\farcs9$) .

As in the case of the continuum emission, no  emission is detected at the
position of the ammonia WHS (see Fig.~\ref{fespectra}). The 3-$\sigma$ upper 
limit from the SMA map is 0.53~\jy\  (2.4~K).

\subsection{SiO}

\begin{figure}
\epsscale{0.45}
\plotone{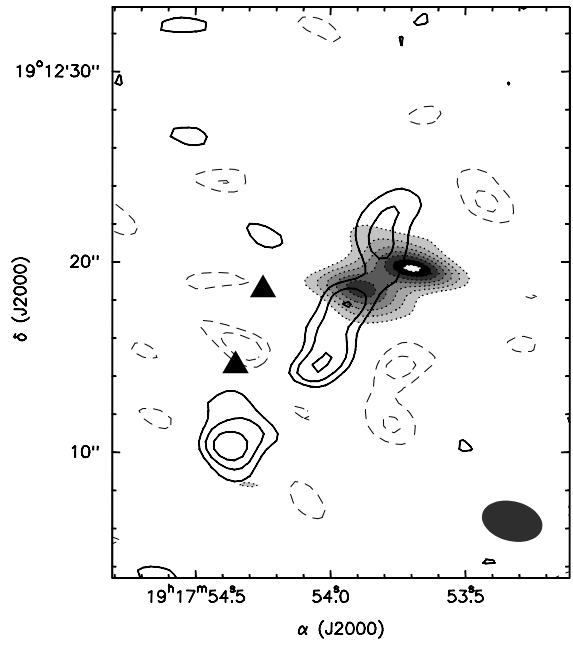}
\caption{
Contour map of the SiO \J{5}{4}\ emission overlaid with the grey scale image
of the 1.35~mm dust.   Contours are  $-3$, $-2$, 2, 3, 4 and 5 times 0.05~\jy,
the rms noise of the map.  The synthesized  beam is shown in the bottom right
corner.
The filled triangles  mark H$_{2}$ knots  \citep{Palacios99}.
 \label{fsio}} 
\end{figure}

The SiO \J{5}{4}\ line is marginally detected towards the  SMA~1 and SMA~2
system (see  Figure~\ref{fsio}).  The emission appears clumpy and  elongated in
the NW-SE direction ($PA\simeq145\arcdeg$), with the strongest emission about
$11''$ SE of SMA~1.  Taking into account the fact that the SiO is a tracer of
molecular shocks, the emission appears relatively ``quiescent'': it is only
slightly  redshifted with  respect to the systemic velocity (the emission is
detected in  the $v_{\rm LSR}=11.31$  and 12.44\kms\ channels) and  the
line-width,   $\Delta v \simeq 2$\kms, is only slightly larger than  that of
the lines tracing the dense circumstellar material such as \form\ and ammonia. 

The emission is apparently more associated with SMA~1, but given the complex
outflow activity in the region \citep{Lee02,Carrasco08} it is difficult to
elucidate with which outflow and powering source it is associated. Nevertheless,
at the SiO velocity and location there is strong CO emission (see channel maps
from Lee et a. 2002), which suggests that the SiO is tracing a region of
interaction between the dense envelope and the outflow \citep{Codella99}.

\subsection{Other molecules}

\begin{figure}
\epsscale{0.55}
\plotone{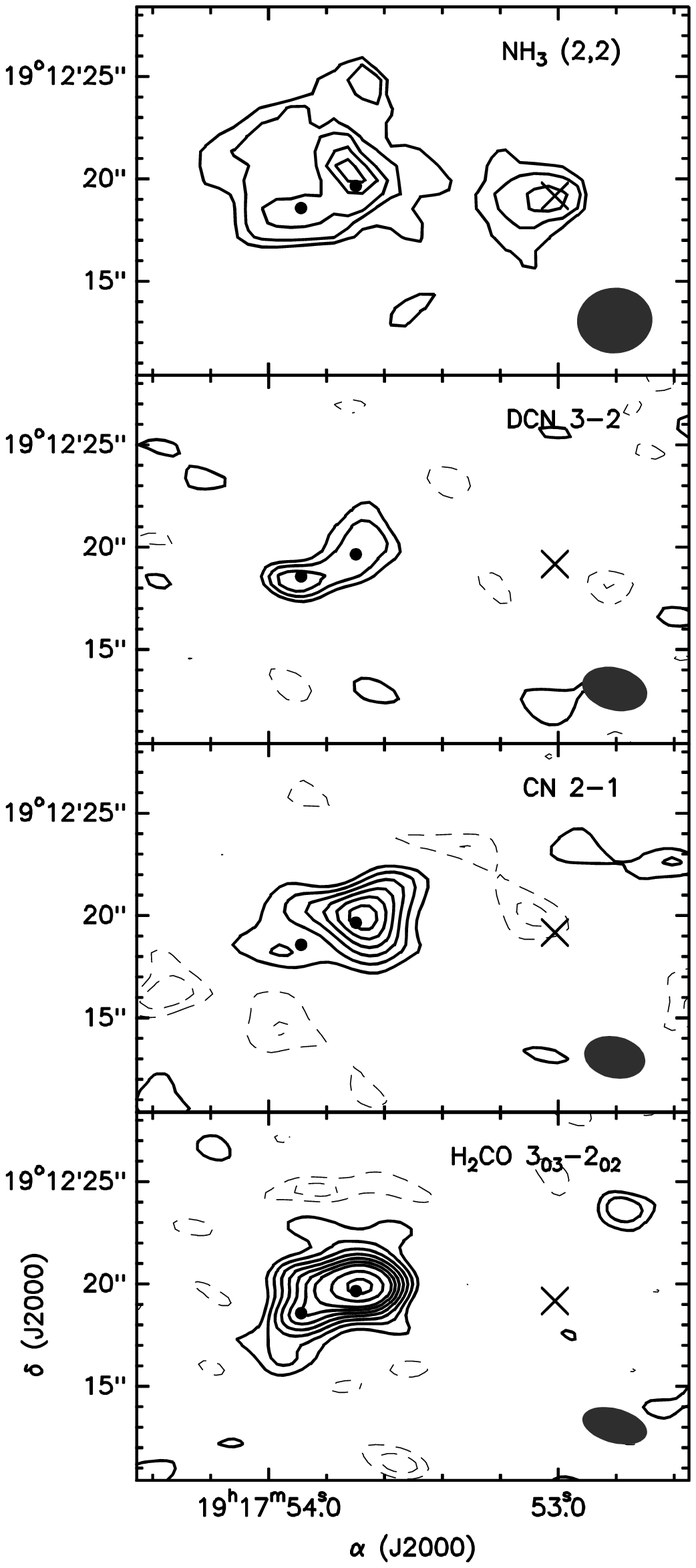}
\caption{
Integrated emission of the \form\ \JK{3}{0,3}{2}{0,2}, CN \J{2}{1}, DCN
\J{3}{2} lines observed with the SMA. For comparison we have also included the
integrated emission of the \nh\ (2,2) line obtained with the VLA by
\citet{Girart97}.  The synthesized  beams are  shown in the lower right corner
of each panel. Contours are $-2$, 2, 3, 4, 5, 6, 7, 8, 10 and 12 times 3.2, 34,
16 and 51~\mjy\ \kms\  for \nh, DCN, CN and \form, respectively. The crosses
show the position of SMA 1 and SMA 2. The titled cross shows the position of the
WHS.
\label{fmol}}
\end{figure}

\begin{figure}
\epsscale{0.7}
\plotone{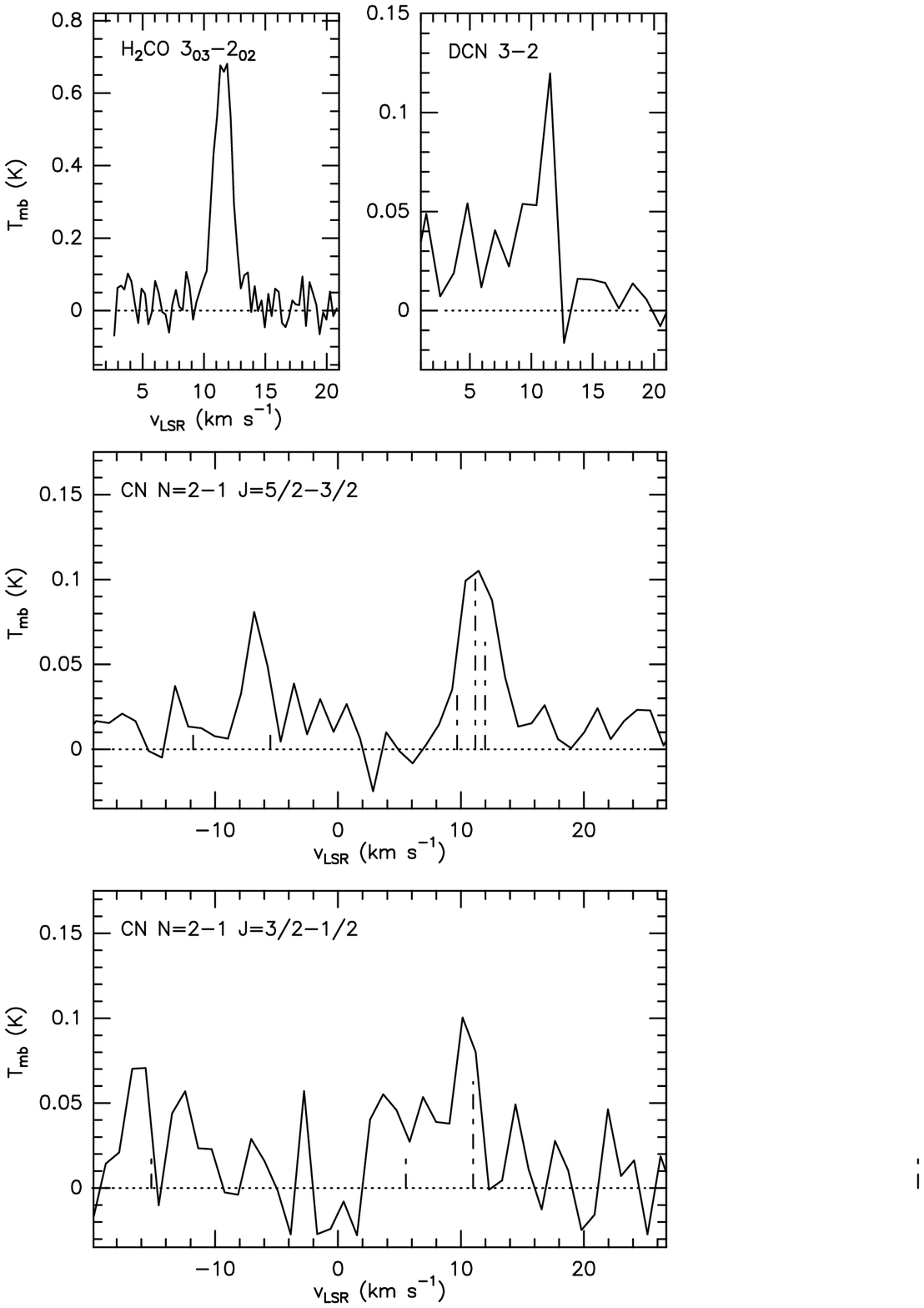}
\caption{
Spectra of the \form\ \JK{3}{0,3}{2}{0,2}, DCN \J{3}{2}, 
CN $N=$2--1, $J=5/2$--$3/2$ and $J=3/2$--$1/2$ obtained 
averaging a region of  $7''\times5''$ around SMA~1 and SMA~2. 
The CN and DCN spectra has a lower velocity resolution ($\simeq 1.0$\kms).
For the CN hyperfine transitions, a vertical dashed line 
indicates the position of the expected hyperfine transition with the height 
proportional to its relative intensity.  
\label{fespe}}
\end{figure}

Several hyperfine transitions of the CN \J{2}{1} rotational transition were
detected (Fig.~\ref{fespe}). The strongest line is a blend of three hyperfine
lines,  the $F$=\J{7/2}{5/2}, \J{5/2}{3/2}\ and \J{3/2}{1/2}\ of the 
$J$=\J{5/2}{3/2} set. The $J$=\J{3/2}{5/2}, $F$=\J{1/2}{3/2}\ hyperfine  line
is also marginally detected. Although there are other hyperfine  transitions
within the bandwidth their relative intensities were too small to  be detected
(see Fig.~\ref{fespe}). The integrated intensity of the detected transitions
(Fig.~\ref{fmol}) shows that the emission is concentrated around VLA~2, being
only weakly detected towards SMA~1. We also marginally detected the DCN
\J{3}{2}\ line toward the two sources  (Figs.~\ref{fmol} and \ref{fespe}).
Despite the poor spectral resolution ($\sim 1$\kms), these two molecules show
emission at the ambient molecular gas velocity with narrow line widths  ($\sim
1$\kms).

\section{Analysis}

\subsection{Model of the dust emission\label{ModelDust}}

\begin{figure}
\epsscale{0.9}
\plotone{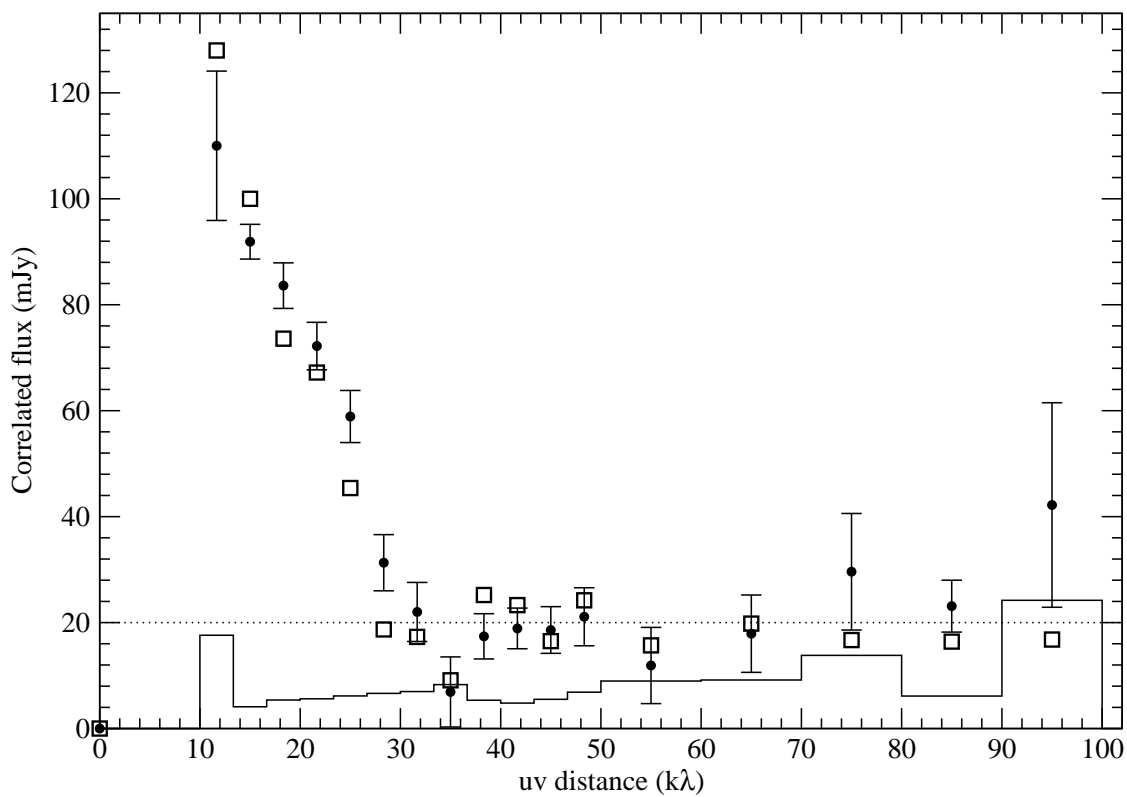}
\caption{ 
SMA correlated flux for the 1.35~mm continuum emission $versus$  uv distance
(in units of k$\lambda$) at the position of  SMA~2.  The correlated flux was
derived by vector averaging the amplitude of the visibilities over annular
bins.  The bins have a width of 3.3~k$\lambda$ for radius lower than 
50~k$\lambda$ and of 10~k$\lambda$ for larger radii.  The solid line  shows the
expected value for the amplitude assuming no signal (i.e., the "zero bias").
The dotted line shows the expected flux for an unresolved source of 20~mJy.
The open squares show the expected correlated for the model $T_0=20$~K  and
$n_0({\rm H_2})=2.4\times10^6$~\cmt\ described in \S~\ref{ModelDust}. 
\label{UVplot}}
\end{figure}

\begin{figure}
\epsscale{0.7}
\plotone{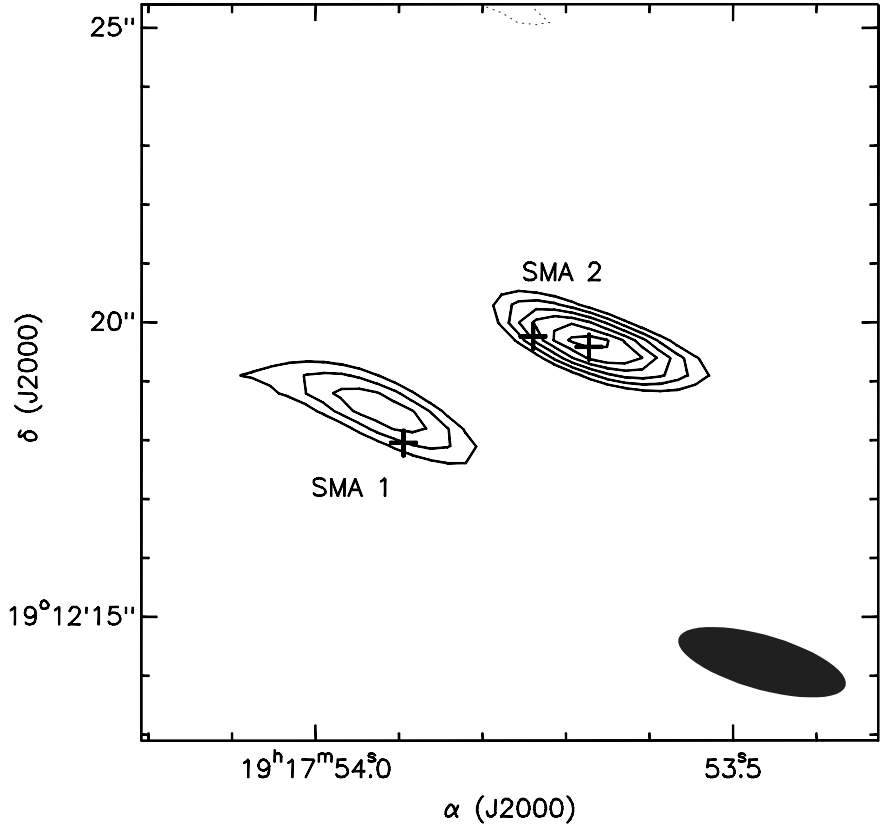}
\caption{
SMA map of the 1.35~mm dust emission obtained using only the visibilities with
a radius larger than 45~k$\lambda$ in the $(u,v)$ plane and applying a Natural
weighting to the visibilities. The contour levels are $-3$, 3, 4, 5, 6 and 7
times the rms noise of the map, 2.4~\mjy. The synthesized beam,
$2\farcs94\times0\farcs92$ with a $PA=74.2\arcdeg$, is  shown in the bottom
right corner. 
The crosses show the position of  VLA~2A (the western cross) and
VLA~2C,  which are associated with SMA 2, and VLA~2D, associated with SMA 1.
\label{fdust4}}
\end{figure}

The dust emission from SMA~1 and SMA~2 are partially resolved with sizes  of
several hundred AU, which suggest that a significant contribution to the 
emission is possibly coming from the infalling  envelopes around the disks. 
Figure~\ref{UVplot} shows clearly that the correlated amplitude in the
visibility data centered on VLA~2 comes mostly from an extended component, 
although there is also a weaker unresolved component, with a flux density of 
$\sim 20$~mJy, that appears to be dominant at visibility radii larger than 
$\sim 40$~k$\lambda$.  To better characterize this apparently compact
component, additional maps were obtained using only visibilities that have a
radius in the $(u,v)$ plane larger than 45~k$\lambda$. The resulting map of the
emission from SMA~1 and SMA~2 appears to be compact.  Both sources are
unresolved with fluxes of $19.7\pm2.4$ and $13.3\pm2.4$~mJy for SMA~2 and
SMA~1, respectively. This is about one third of the total flux density detected
by the SMA for each source. A Gaussian fit to SMA~2 using the task IMFIT of
AIPS yields an upper limit of the diameter of $\simeq 1\farcs2$: The other
source, SMA~1, is too weak to do  the same fitting. In any case, the compact
dusty components seem to arise from a structure with a radius of $\la 180$~AU.
Without proper modeling it is not possible to elucidate whether this compact
component traces the disk or the inner part of the envelope. Interestingly, the
peaks of the two components are slightly shifted ($\sim0\farcs4$) with respect
to the peak position given in Table~\ref{tdust}. This could be produced by a
somewhat asymmetrical distribution of the gas and dust in the envelope, which
will not be a surprise because of the multiplicity detected with the VLA at the
center of SMA~2. The compact component of SMA~2 coincides well with the
centimeter source VLA~2A detected by \citet{Carrasco08}
(the position of the dust is
$\alpha$(J2000)$=19^h17^m53\fs666$ and
$\delta$(J2000)$=19\arcdeg12'19\farcs63$). 
The peak of the SMA~1 compact component 
($\alpha$(J2000)$=19^h17^m53\fs920$ and
$\delta$(J2000)$=19\arcdeg12'18\farcs52$) is displaced to the northwest with
respect to VLA~2D by $0\farcs67$ (see Fig.~\ref{fdust4}). Given the weak
detection of this source at both 1.35 and 7 mm (5 and 4-$\sigma$ respectively)
further more sensitive observations are needed to confirm this displacement.

\begin{figure}
\epsscale{0.8}
\plotone{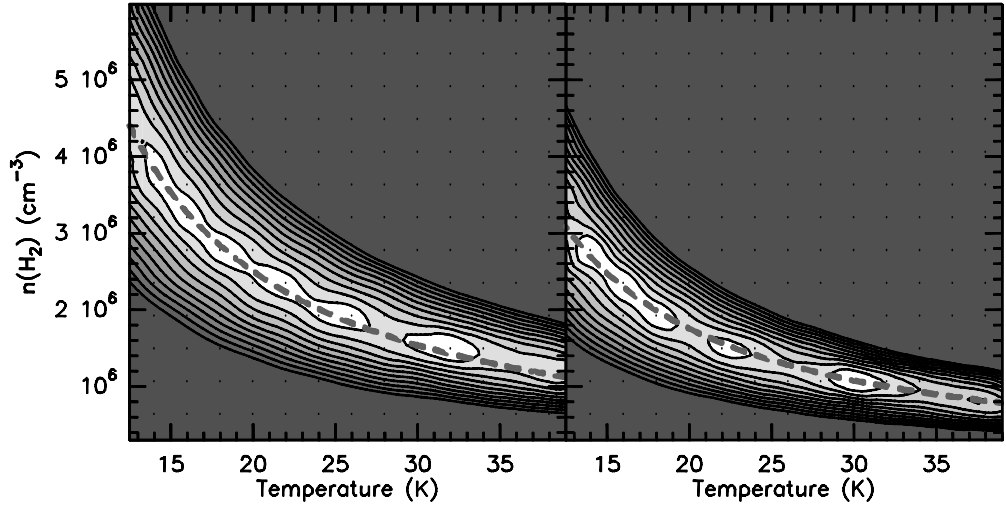}
\caption{
Plots shows the locus of best solutions represented by the function of the 
{\em rms} of the residual map derived by subtracting the dust 
model for an inner radius of 30 ({\em left panel}) an 100~AU ({\em right 
panel}) to the SMA visibility data of the 1.3~mm continuum  emission. Levels 
are in steps of 3 times the {\em rms} of the original SMA 
map, 1.5~\mjy. The dashed grey line shows the line of approximately set of best solutions represented as the function  
$n{\rm (H_2)}=1.5\times10^6 \, [T/30 \, {\rm K}]^{1.2}$~\cmt\ (left panel) and
$n{\rm (H_2)}=1.1\times10^6 \, [T/30 \, {\rm K}]^{1.2}$~\cmt\ (right panel).
\label{fresidumodel}}
\end{figure}

In order to characterize the properties of the double cores we modeled the dust
emission assuming, for simplicity, that it arises from two independent 
optically--thin twin circumstellar envelopes with a radial density and 
temperature profiles at the position of the dusty sources detected with the
SMA. The model  integrates the dust emission assuming spherical symmetry from
an  inner radius, $R_{\rm inner}$, to an outer radius of $1.5\times10^4$~AU
(this value is somewhat arbitrary and does not affect the fit because the
emission  at these scales is filtered out by the SMA).  We do not include the
possible contribution from the circumstellar disks, since it is possible that
at  this stage the disks are small, with a radius of $\la 30$~AU, 
\citep{Rodriguez05}.  The three possible protostars associated with SMA~2  are
separated by a projected distance of $\simeq0\farcs9$ or 270~AU 
\citep{Carrasco08}, below the angular resolution of our observations. If  these
sources are truly nearby protostars, it is possible that there is a  cavity in
the envelope between these sources. To check whether there is a  significant
cavity, we used different radii of 30, 100 and 180~AU.  We  adopted the density
profile expected for an infalling envelope,  $n(r) \propto r^{-1.5}$, up to an
infall radius, $R_{\rm infall}$. Beyond  $R_{\rm infall}$ we used a $r^{-2}$
density profile.  In order to minimize the free parameters in the model, we
adopted a $R_{\rm infall}=1000$~AU,  the value found by \citet{Shirley02}.  For
a dusty cloud  heated by an internal source, the temperature radial profile can
be characterized as $T \propto r^{-2/(4+\beta)}$ \citep{Kenyon93}, where
$\beta$  is the dust emissivity spectral index.  The value used is $\beta=1.5$,
which  was derived from 850 and 450~$\mu$m SCUBA observations
\citep{Estalella03}.  We adopted an 1.35~mm dust opacity of $\kappa
\simeq0.9$~cm$^2$, which should  be adequate for dusty particles  with ice
mantles at densities of about  10$^6$~\cmt\ according to \citet{Ossenkopf94}.  
Additional free parameters were the density $n_0$ and temperature $T_0$ at 
a radius of 1,000~AU. Intensity profiles were obtained for each set of 
$R_{\rm inner}$, $n_0$ and $T_0$ and converted to a 2-D image map. 
These modeled maps were multiplied by the primary beam response of the 
SMA antennae, assumed to be $56''$ at 1.35~mm. Two sets of visibility data 
were generated: one obtained by subtracting the modeled map to the 
visibility continuum data of the SMA observations. The other set was obtained 
by  replacing the values of observed visibilities with the values expected 
from the  modeled map. Maps of the residual and of the model were obtained 
from the  visibilities using the same parameters as the SMA maps shown here. 
The {\em rms} and {\em bias} was finally computed for these residual maps. 
The bias is defined as the absolute flux density of the residual on a 
given region (the region where the emission is detected).  The models with 
an inner radius of 180~AU cannot reproduce well the SMA data, which suggests 
that if there is an inner cavity (produced by the presence of a multiple 
protostellar system), it is not big enough to affect the simple model used here. 
With respect to the solutions with an inner radius of 30 and 100~AU, there 
is, for both, a family of possible solutions in the $T_0$--$n_0$ plane. 
The plot of the {\em rms} for these two inner radii in the range of densities 
and temperatures computed are shown in Figure~\ref{fresidumodel}. 
The best set of solutions can be expressed as 
$n_0=$ $1.5\times10^6 \, [T_0/30 \, {\rm K}]^{-1.2}$~\cmt\ and  
$1.1\times10^6 \, [T_0/30 \, {\rm K}]^{-1.2}$~\cmt\ for
the 30 and 100~AU inner radius, respectively. 

\begin{figure}
\epsscale{0.8}
\plotone{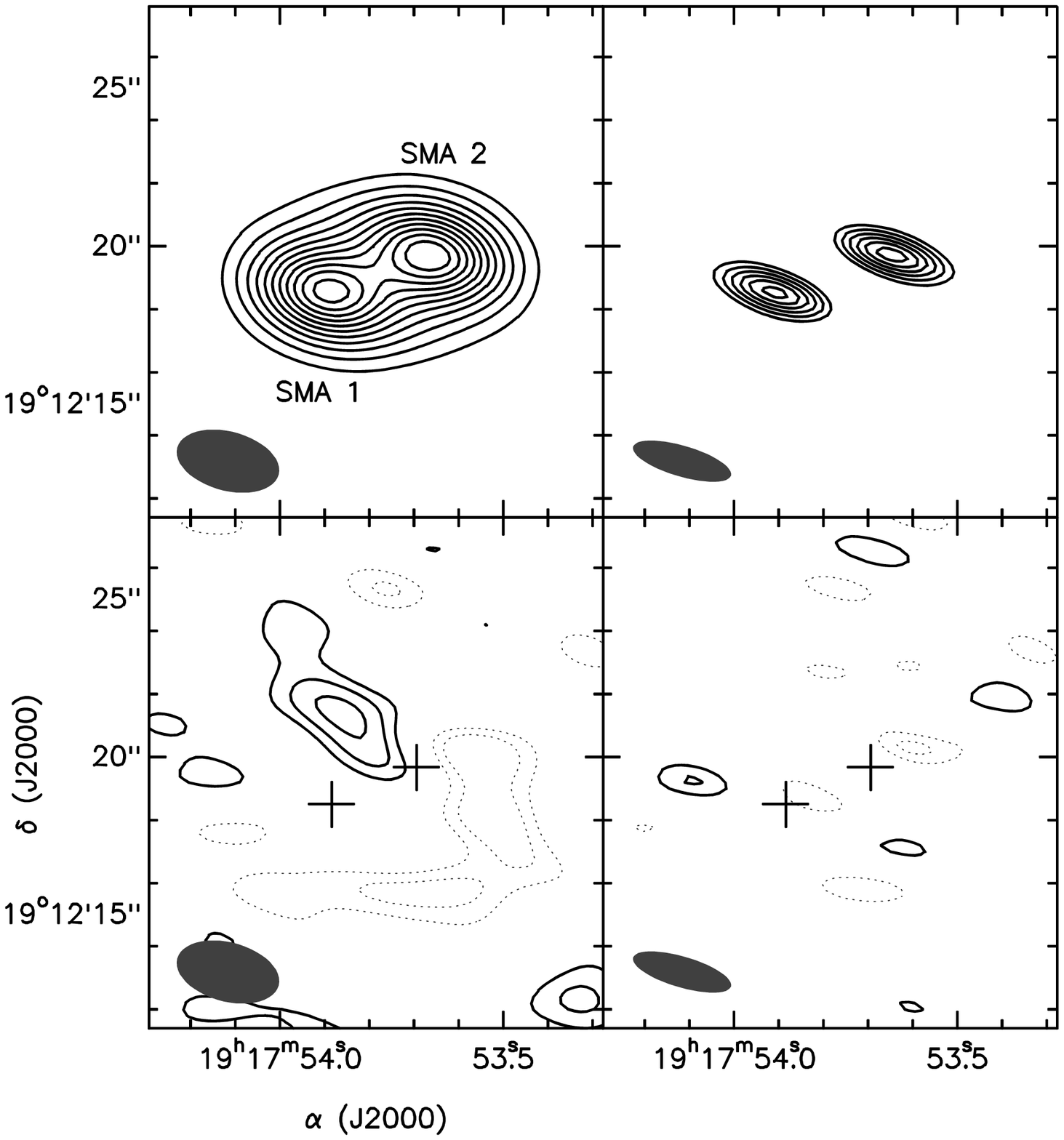}
\caption{
SMA synthetics maps for the dust model $R_{\rm inner}=30$~AU, $T_0=20$~K 
and $n_0=2.4\times10^6$~\cmt\ described in
\S~\ref{ModelDust}. 
{\em Left panels}: the model (top) and residual (bottom) dust 
maps obtained using all the visibilities observed with the SMA
and using robust of 1. 
Contours levels for the model map are are $-3$, 3, 5, 7, \nodata,
21 times 1.5~\mjy, the $rms$ of the map.
Contour levels for the residual map are $-3$, $-2$, 2, 3, and
4 times 1.5~\mjy,
{\em Right panels}: the model (top) and residual (bottom) dust maps 
obtained using visibilities with a radius larger than 45~k$\lambda$
and Natural weighting (as in Fig.~\ref{fdust4}).
Contour levels are $-3$, $-2$, 2, 3, 4, 5, 6 and 7 times 2.4~\mjy. 
The synthesized  beams are shown in the lower left corner of
each panel. The crosses show the position of SMA~1 and SMA~2.
\label{fmapmodel}}
\end{figure}

The VLA ammonia maps from \citet{Girart97}  can be used to further constrain 
the model. At the positions of SMA~1 and SMA~2 the kinetic temperature
estimated  from the \nh\ (1,1) and (2,2) maps is $\simeq25$~K (see
\S~\ref{ResultDust}).  Since the beam of the VLA maps is $\simeq3\farcs5$, it
is reasonable to adopt  the value of $1\farcs75$ or 520~AU, as the radius at
which the gas is at 25~K.  This implies that the temperature profile of the
dust  can be written as  $T = 20 \, (r/1000\, {\rm AU})^{-0.35}$~K. The
previous two equations  that give the best set of solutions yield a density
profile of $n({\rm H_2}) = $ $2.4 \times 10^6 (r/1000\, {\rm AU})^{-1.5}$~\cmt\
for  $R_{\rm inner}=30$~AU, and of $1.8 \times 10^6 (r/1000\, {\rm
AU})^{-1.5}$~\cmt\  for $R_{\rm inner}=100$~AU. Figure~\ref{UVplot} shows that
the correlated  flux in the visibility domain for the model with $R_{\rm
inner}=30$~AU,  $T_0=20$~K and $n_0=2.4\times10^6$~\cmt\ fits reasonably well
the SMA data.  The synthetic maps from the model and from the residuals for
this particular  solution also show the remarkable similitude with the L723
dusty binary system  (see Fig.~\ref{fmapmodel}). Indeed, from  the residual map
obtained using only the longest baselines (see right panels of 
Fig.~\ref{fmapmodel}) the model of the binary envelope can account within  the
$rms$ level of the map the compact, apparently unresolved, component of the two
dusty sources, without the need of invoking the presence of an accretion disk
(see the discussion section). On the other hand, the residual map done using
all the visibilities shows some significant residuals (at a $\sim4$-$\sigma$
level): there are some positive residuals northeast of SMA~2  and extended
negatives on the southern side of the dust emission from both SMA~1  and SMA~2.
These residuals could be due to the departure from the spherical  symmetry of
the data.

We can also compare the density we found with the value derived by 
\citet{Dartois05} by modeling the spectral energy distribution of L723.  They
estimated that at 100 AU the density is $2.2\times10^7$\cmt, which is  a factor
2.6 to 3.4 lower than what we derive for $T_0=20$~K. This difference could be
due to the different dust opacity and the different density profile used.

The derived density distribution can be used to estimate the mass of the two 
envelopes. The equivalent radius of the SMA shortest baselines in the 
visibility domain ($\simeq12$~k$\lambda$) is $3\farcs75$ (1125~AU).   The 
combined mass for the two envelope within a radius of 1125~AU is  in the
0.14--0.36~\mo\ range for a temperature range of $T_0=20$--30~K and  of $R_{\rm
inner}=30$--100~AU. These values are compatible with the mass  estimated in
Table~\ref{tmass}.

\begin{figure}
\epsscale{1.0}
\plotone{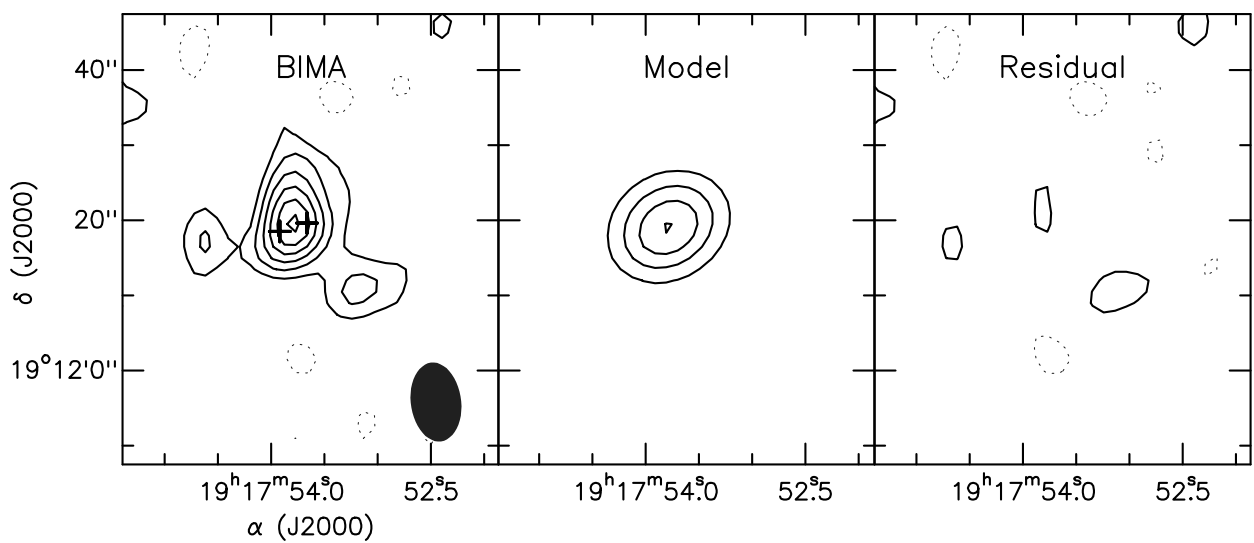}
\caption{
{\em Left panel}: BIMA map of the 3.2~mm continuum emission obtained 
with the C configuration in September 2003. The synthesized beam,
$10\farcs6\times6\farcs7$ and $PA=-7\arcdeg$, is shown in the bottom 
right corner. Contours are $-2$, 2, 3, 4, 5, 6 and 7 times 2.8\mjy,
the rms noise of the map. The crosses show the position of SMA~1 and SMA~2.
{\em Middle panel}: 
BIMA synthetic map at 3.2~mm continuum at the C configuration
for the dust model $T_0=20$~K and $n({\rm H_2})_0=2.4\times10^6$~\cmt\ 
described in \S~\ref{ModelDust}. Contours are the same as in the previous panel.
{\em Right panels}: BIMA residual map obtained from the previous two maps.
\label{fbimamodel}}
\end{figure}

As an additional test for the model, we have used a 3.2~mm continuum map
obtained with the BIMA array. These observations were carried out in  September
2003 in the C configuration. The continuum was observed simultaneously with the
N$_2$H$^+$ 1--0 and CH$_3$OH $2_0$--$1_0$~A and $2_{-1}$--$1_{-1}$~E  lines
(Masqu\'e, Girart \& Estalella, in preparation).  
The dust model was extrapolated to this wavelength, and the standard position 
of the BIMA antennas at the C configuration were used to create a synthetic 
BIMA map using the same procedure as the one described for the SMA. The
synthetic maps were obtained for the $R_{\rm inner}=30$~AU, $T_0=20$~K and 
$n_0=2.4\times10^6$~\cmt\ model.  Figure~\ref{fbimamodel} shows  that the model
also reproduces reasonably well the BIMA map.

\subsection{Modeling the H$_2$CO emission} \label{smodelform}

Once the dust emission was successfully modeled, we modeled the \form\
emission using the one dimensional version of Ratran  \citep{Hogerheijde00}.
This is a Monte Carlo code that calculates the radiative  transfer and
excitation of molecular lines. The code is formulated from the  viewpoint of
cells rather than photons, which allows the separation of local  and external
contributions of the radiation field. This gives an accurate and fast 
performance even for high opacities \citep{Hogerheijde00}. The \form\ 
collisional rates used were derived by \citet{Green91} and were downloaded 
from the Leiden Atomic and Molecular Database \citep{Schoier05}.

We modeled the emission as arising from two cores with the same density and
temperature profile, as suggested by the dust modeling.  Since there is
a family of possible solutions, we fixed
the temperature and density profiles derived in the previous section that
agrees with the ammonia rotational temperature: 
$T = 20 \, (r/1000\, {\rm AU})^{-0.35}$~K and 
$n({\rm H_2}) = 2.4 \times 10^6 (r/1000\, {\rm AU})^{-1.5}$~\cmt\
(the solution for $R_{infall}=30$~AU).

\begin{figure}
\epsscale{0.9}
\plotone{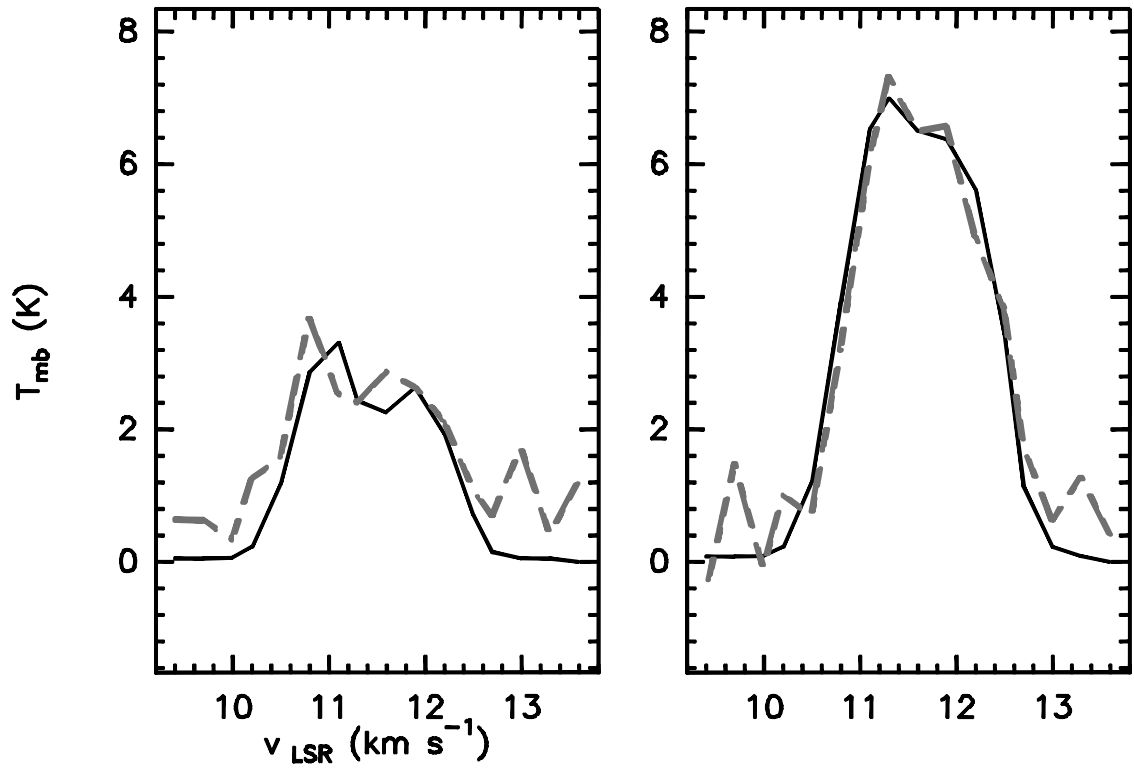}
\caption{
Synthetic (solid line) and observed (dashed grey line) \form\
\JK{3}{0,3}{2}{0,2}\ spectra at SMA~1 (left panel) and SMA~2 (right panel).
The model parameters for the synthetic spectra are the same as in
Figure~\ref{fmodel1}.
\label{fmodel2}}
\end{figure}

The kinematics is not well resolved with our observations, so we assume  that
the gas is in free-fall collapse.  This is marginally suggested by the line
profile at the position of SMA~1 and SMA~2 (see Fig.~\ref{fmodel2} and
\S~\ref{Stwin}). We adopted a velocity field for the free-falling gas of  
$v_{\rm infall} = 0.5 \, (r/1000\, {\rm AU})^{-0.5}$\kms\  and an intrinsic
linewidth of 0.6\kms.  Using these values the modeled line width was similar to
the  \form\ values measured with the SMA and, in any case, the use of these
values should not affect critically the derivation of the \form\ properties
\citep{Jorgensen04}. 

To run RATRAN we adopted a fixed \form\ abundance within a shell with an inner 
and outer radius $R_{in}$ and $R_{out}$, respectively, with the abundance being 
zero outside the shell.  We ran a series of models with Ratran using a range of 
values for $R_{in}$ and $R_{out}$ and for the p--\form\ abundance, 
$X[$p--\form$]$. Since the \form\ emission is significantly different between 
the two envelopes, we made synthesized maps using different sets of ($R_{in}$, 
$R_{out}$, $X[$p--\form$]$) values for  SMA~1 and SMA~2. Synthesized maps  with 
16 channels and with a spectral resolution of 0.28\kms\ were generated. We took 
into account that the systemic velocity of the \form\ between the two cores 
differs by $\simeq0.2$\kms\ (see Table~\ref{tform}). The maps were multiplied by 
the primary beam of the SMA and then converted to visibilities. Maps of the 
model and of the residuals (generated by subtracting the model to the data in 
the visibility domain) were made in the same fashion as for the SMA dust 
emission maps (see \S~\ref{ModelDust}).  

We explored values within the range of $8\times10^{-11}$ to $1\times10^{-8}$
for $X[$p--\form$]$, 30 to 240~AU for $R_{in}$ and 300 to $10^4$~AU for
$R_{out}$. In order to find the best set of solutions, we computed separately
the $\chi^2$ parameter for the spectra at the SMA~1 and SMA~2 positions and for
the total flux density within a region of $11''\times8''$ centered around the
twin envelopes. Table~\ref{tRatran} shows the range of the best fits found. The
range of parameters is better constrained for SMA~2 than for SMA~1. This is
due to its higher intensity and its apparently more compact size. Indeed, the
model yields solutions that imply that the \form\ around SMA~2 should arise
from the inner region of the envelope. More specifically, the outer radius
cannot be significantly larger than 600 AU, but not lower than 300~AU
(otherwise the emission would be unresolved, which is not the case). The SMA~2
abundance is also relatively well constrained,
$X[$p--\form$]\simeq(3$--$10)\times10^{-10}$. For SMA~1, the best set of
solution yields \form\ abundances in the range 
$X[$p--\form$]\simeq(8$--$30)\times10^{-11}$, but with values always
lower than for SMA~2: the best solutions are found when SMA~2 abundances are a
factor of 3 to 10 higher than for SMA~1.  Another difference with respect to
SMA~2 is that the  best models for SMA~1 have a larger outer radius than for
SMA~2, $R_{out}\simeq 600$--5000 AU. 

\begin{figure}
\epsscale{0.55}
\plotone{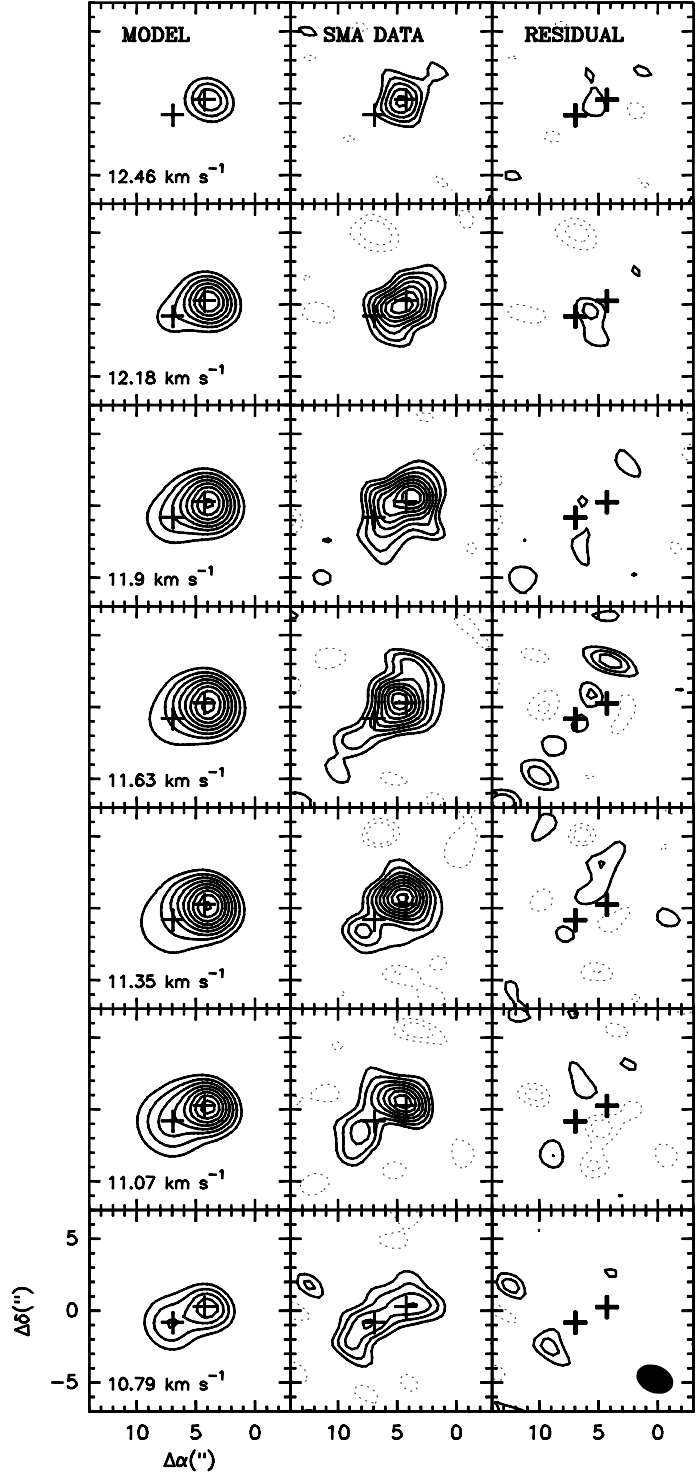}
\caption{
Channel maps of one of the best models, the SMA data and the residual
(SMA$-$model) for the \form\ \JK{3}{0,3}{2}{0,2}\ emission. Contours are $-3$,
$-2$, 2, 3, 4, 5, 6, 7 and 8 $\times$ 0.17~\mjy. The synthesized  beam is shown
in the lower right corner of the lower right panel. The crosses show the
position of the dust sources SMA~1 and SMA~2. The parameters for this model are
$X[$p--\form$]=3\times10^{-9}$,  $R_{in}=240$~AU and $R_{out}=600$~AU for
SMA~2,  and  $X[$p--\form$]=1\times10^{-10}$, $R_{in}=60$~AU and
$R_{out}=5000$~AU for SMA~1. 
The \vlsr\ velocity of the channels is shown in the lower left corner
of the right panels.
\label{fmodel1}}
\end{figure}

Figure~\ref{fmodel1} shows the synthesized model, the SMA data and the residual
for one of the best solutions found (see the figure caption for details). 
Figure~\ref{fmodel2} shows the spectra toward SMA~1 and SMA~2 for the same
model and for the SMA data.  These figures show that the models fit the
observations reasonably well, although the residuals appear up to about a
4-$\sigma$ level. In spite of the good match between the synthesized and
observed spectra for the two sources, the model reproduces better the
morphology of SMA~2 than that of SMA~1. The morphology mismatch between SMA~1 and
the synthesized map occurs for all the models. This suggest that the emission,
specially from SMA~1, departs from the idealized spherical model used with the
assumption that the \form\ arises only within a shell. Indeed, the
position--velocity cuts in two orthogonal directions centered on SMA~1 (see
Fig.~\ref{fpv}) show a double peak in one of them and what roughly appear to
partially be a ``donut'', centered on SMA~1, in the other cut (partially
because of the overlap with SMA~2 in the NW side). This suggests that the
\form\ emission may not come from the shell of a spherical envelope but from a
contracting and somewhat flattened shell or torus. For example, such a case was
more clearly observed in the CS emission associated with the dense core ahead
of HH~80N \citep{Girart01}. The marginal evidence that the \form\ arises from a
torus supports the idea that the emission does not come from  a spherical
configuration. However, despite the limitations of the adopted approach,
given the limited signal-to-noise ratio obtained and the use of only one
transition, the results from the modeling should be regarded as a good first
approximation to the properties of \form\ in these two cores.

The derived p--\form\ abundances are compatible with the value found by
\citet{Jorgensen05},  $9.6\times10^{-10}$,  from single-dish observations, and
are also within the range of values found for the so--called outer part of the
circumstellar envelopes associated with Class 0 sources
\citep{Jorgensen05,Maret04}, which are defined as the region where temperatures
are lower than $\simeq100$~K. In the inner region of the envelopes, where the
temperature is $\ga 100$~K, the icy grain mantles evaporates and molecular
abundances are expected to increase significantly: the \form\ reaches abundances
between $10^{-8}$ and $10^{-6}$  \citep{Jorgensen05,Maret04}. Our model did not
take into account this warm component. Nevertheless, the reasonable solutions
found in this paper imply that the total amount of \form\ in the inner warm
region is not significant, despite the very high \form\ abundance expected. This
can be explained by the fact that for L723, the warm region is very small  due to
its low total bolometric luminosity, 3.4~L$_{\odot}$ (the temperature of 100~K is
reached at a radius of 10~AU).

\section{Discussion}

\subsection{The Twin Cores\label{Stwin}}

The dust emission reveals two similar substructures, with SMA~2 being only
slightly more massive than SMA~1 (Table~\ref{tdust}). In fact, the modeling of
the emission suggest that within the observational uncertainties (sensitivity
and angular resolution) the dust properties of the two cores can be accounted
as arising from ``identical'' twin cores.  However, these twin cores have
different molecular properties. The CN \J{2}{1} and \form\ \JK{3}{03}{2}{02}
emission is stronger in SMA~2 than in SMA~1. In fact, the modeling carried out
in \S~\ref{smodelform} shows that the \form\ associated with SMA~2 has a higher
molecular abundance (by a factor 3--10) and arises from a more compact region
than in SMA~1. In both cores, the line width of the \form\ lines are clearly
larger (by up to 50\%) than those of \nh. In addition, the \form\
\JK{3}{03}{2}{02} emission is more compact than the \nh\ emission, as can be
clearly seen in Fig.~\ref{fmol}. These two features taken together are a good
indication of the presence of radial velocity gradients, with increasing
velocity with decreasing radius to protostars. Both infall and rotation
velocities are expected to have this behavior. The presence of a blueshifted
peak slightky higher than the redshifted one at the position of the two cores
(Fig.~\ref{fmodel2}) is an additional marginal evidence of infall motions. 

Previous observations find clear star formation signatures associated
with the SMA~2 core. First the presence of a well studied thermal radio jet,
VLA~2A, pinpoint the position of a protostar that is powering the east--west
molecular outflow \citep{Anglada96,Carrasco08}. VLA high angular resolution
observations reveal the presence of a water maser, and 7~mm and 3.6~cm
sources (VLA 2B and 2C) $\sim0\farcs5$ around VLA~2A
\citep{Girart97,Carrasco08}. \citet{Carrasco08} interpret these as a young
stellar system, which is forming in the 150~AU  vicinity (in projected
distance) around VLA 2A. In addition, they find that VLA~2B may be powering
the north-south molecular outflow. This stellar system is being formed inside
the SMA~2 core. In contrast, there are not known signs of outflow activity
towards SMA~1, yet. There is only marginal dust emission detected at 7~mm,
labeled as VLA 2D \citep{Carrasco08}.

In conclusion, our study and analysis of SMA~1 and SMA~2 show that they have
almost identical properties (mass, and density and temperature distribution)
and that there is marginal evidence of infall motions in both. However, the
\form\ abundance towards SMA~2 is significantly higher than in SMA~1. Since
SMA~2 shows clear outflow activity, in contrast with SMA~1, this suggests that
SMA~2 is more evolved (due to a higher infall rate or because it is older). In
any case, they are not ``identical'' twin cores but ``fraternal'' twin cores.

\subsection{The Mass Reservoir around the Class 0 protostellar system}

The total mass derived from the SMA observations is less than 10\% of the whole
dense core, derived from SCUBA JCMT observations at 850 and 450 $\mu$m
\citep{Estalella03} or from VLA \nh\ observations \citep{Girart97}. That is,
the mass accreting to the protostars (or to the disks) at scales of $\la
1000$~AU is only a small part of the dense core. In addition, from the modeling
we found that the dust emission detected by the SMA at 1.35~mm seems to arise
from the  dense twin envelopes, SMA 1 and SMA 2, and within the limits of our
sensitivity there is no need to invoke to the presence of protostellar disks.
That  means that the contribution from the disks at 1.35~mm should not be
higher  than $\simeq 8$~mJy (the 5--$\sigma$ value of the continuum residual
map  of Fig.~\ref{fmapmodel}). Assuming that the disk has a temperature of 50
K,  the upper limit of the disk mass is $6\times10^{-3}$~M$_{\odot}$. This mass
would be higher if the dust emission from the disk is not optically thin, but
it would be lower if the temperature is higher. In any case, this suggests that
the disk mass is less than 10\% of the mass of each of the twin cores.  Thus,
the large scale (0.1~pc) molecular envelope traced by \nh\ or SCUBA contain
most of the mass or, in other words, it is still the major reservoir of mass to
keep the star formation going on.  Therefore, L723 seems to be in a very early
phase of star formation.

It is clear that fragmentation is occurring at different levels.  The $0.1$~pc
scale molecular envelope has fragmented at the center and at scales of
$\simeq1000$~AU  into two cores, SMA 1 and SMA 2.   At the same time, at least
one of these cores, SMA~2, has undergone additional fragmentation at scales of
$\simeq150~$AU, forming a multiple stellar system. However, we note that within
the uncertainties,  the SMA 2 can be modeled a single envelope down to a
radius of 30 AU.  It is interesting to note
that whatever is the fragmentation process that is acting in the different
scales (turbulence, the initial magnetic field and angular momentum
configuration) is occurring  within a 0.1~pc dense envelope that appears to be
very quiescent at scales of few thousands AU outside its center according to
the \nh\ data \citep{Girart97}: line widths of only $\la0.5$\kms\  and small
velocity gradients, of 3\kms~pc$^{-1}$  .

\section{Summary}

We present continuum and line 1.35~mm high angular resolution observations
($\simeq3''$) carried out with the SMA toward the low--mass protostellar region
in the L723 dark cloud. We detected emission of the dust, the \form\ 
\JK{3}{0,3}{2}{0,2}, DCN \J{3}{2} and SiO \J{5}{4}\ transitions, as well as
a few lines of the CN \J{2}{1} transition. We performed a radiative transfer
analysis of the dust emission as well as for the \form\ line in order to
constrain the physical and chemical conditions of the two cores. The main
results of the paper can be summarized as:
\begin{enumerate}
\item The dust emission arises from two similar cores, SMA~1 and SMA~2,
separated by $2\farcs9$ (880 AU in projected distance) and with masses of 0.06
and 0.08~M$_{\odot}$, respectively. They are partially resolved, with
deconvolved sizes of $\sim$500~AU. SMA~2 is associated with VLA~2 and water
maser emission. We modeled the dust emission from the two cores assuming that
the cores have a density and temperature radial distribution of the type:  
$n(r) \propto r^{-1.5}$ and $T \propto r^{-0.35}$.  We found that within the
uncertainty achieved, the two cores can be considered  almost ``identical''
twins. The best set of solutions are those with a density and temperature that
match the following expression:  $n_0{\rm (H_2)}=1.45\times10^6 \, [T_0/30 \,
{\rm K}]^{1.2}$~\cmt\  ($n_0$ and $T_0$ are the density and temperature at a
radius of 1000~AU). The model shows that all the emission detected can arise
from the cores, with possibly no evidence of contribution from the accretion 
disk, which yields an upper limit of the disk mass of
$\sim6\times10^{-3}$~M$_{\odot}$.

\item The \form\ emission is concentrated around the two mm continuum sources,
SMA~1  and SMA~2, although the emission is brighter toward SMA~2.  The
kinematics of the two sources is difficult to disentangle because their
separation  is similar to the angular resolution. Nevertheless, there is
marginal evidence of infall motions in both sources (double peak line, with
brighter blueshifted component), of a velocity gradient around SMA~2 indicative
of rotation, and that the \form\ in SMA~1 arises  from a contracting and
somewhat flattened structure. The linewidth of \form\ has values in the 
1.2--1.7 \kms\ range, which is $\sim$50\% larger than the values measures for
the  \nh\ lines by \citet{Girart97}.  The \form\ emission was modeled adopting
the density and temperature profile constrained by the dust emission. We used
RATRAN assuming a fixed p--\form\ abundance within a shell with a inner and
outer radius $R_{in}$ and $R_{out}$, respectively, being the abundance zero
outside the shell.   Reasonable models were found for a p--\form\ abundance
range of $(3$--$10)\times10^{-10}$ and (8--$30)\times10^{-11}$ for SMA~2
and SMA~1, respectively, but with the SMA~1 abundance lower than for SMA~2 by 
a factor 3 to 10.   The best models are those that have a more compact emission
in SMA~2 ($R_{out}\simeq 300$--600 AU) than in SMA~1 ($R_{out}\simeq 600$--5000
AU).  The p--\form\ abundances found  are compatible with the value of the
outer part of the circumstellar envelopes associated with Class 0 sources. The
total amount of \form\ in the inner warm region of the circumstellar envelopes
(where $T\ga 100$~K) is not significant, despite the very high \form\ abundance
expected (e.g.; Maret et al. 2004). This can be explained by the fact that for
L723, the warm region is very small (a radius of $\la 10$~AU) due to its low
total bolometric luminosity, 3.4~L$_{\odot}$.

\item The DCN and CN lines arise from the same region than the dust and
\form. The CN emission peaks clearly toward SMA~2, whereas the DCN shows
similar emission in SMA~1 and in SMA~2. 

\item SiO is detected marginally at slightly redshifted velocities
around SMA~1 and SMA~2, and is possibly tracing a region of interaction between
the dense envelope and the outflow. 

\item The Western Hot Spot (WHS), detected by \citet{Girart97} from \nh\
observations as a spot of local heating, is not detected by the SMA (line and
continuum) observations. The non--detection of the 1.35~mm emission yields an
upper limit of the WHS mass of $\simeq$0.01~M$_{\odot}$.

\item  The above results suggest that although SMA~1 and SMA~2 cores  have
almost identical physical properties (density and temperature distribution),
they are rather ``fraternal''  twin cores, with SMA~2 being in a more evolved
stage. Thus, SMA~2 is harboring an active multiple low--mass protostellar system
(VLA~2A, 2B and 2C), powering at least one molecular outflow
\citep{Carrasco08}. In contrast, there are not signs of outflow activity
towards SMA~1.  

\item  The large scale, 0.1~pc,  molecular envelope (traced by \nh\ or SCUBA)
is the major reservoir of mass to keep the star formation going on (it contains
about $\ga 90$\% of the mass), which suggests that L723 is still in a very
early phase of star formation.

\item  Fragmentation is occurring at different levels at the center of the L723
dense molecular cloud: at scales of $\simeq1000$~AU  the cloud has fragmented
into two cores, SMA 1 and SMA 2, and SMA~2 has undergone additional
fragmentation at scales of $\simeq150~$AU, forming a multiple stellar system
\citep{Carrasco08}. Whatever is the fragmentation process that is acting at the
different scales (turbulence, the initial magnetic field and angular momentum
configuration) it is occurring  within a 0.1~pc dense envelope that appears to
be at larger scales. 

\end{enumerate}

\acknowledgments
JMG is grateful to the SMA staff at Hilo for the support during the visit to
the SMA.  We would like to thank the anonymous referee for the valuable
comments. JMG acknowledges support from the DURSI  (Generalitat de  Catalunya)
grant BE-2004-00370.  JMG and RE are  supported by MICINN grant
AYA2005-05823-C03 (co-funded with FEDER funds).  


\clearpage

%
%

\clearpage

\clearpage

\clearpage

\clearpage

\clearpage 

\clearpage 

\clearpage 

\clearpage 

\clearpage 

%

%

\clearpage 

\clearpage

%
%

\begin{deluxetable}{lrccccc}
\tablecaption{Parameters  of the SMA observations\label{tobserva}}
\tablewidth{0pt}
\tablehead{
\multicolumn{2}{c}{}&
\multicolumn{2}{c}{Synthesized Beam}&
\colhead{} &
\colhead{Spectral} &
\colhead{$rms$}
\\
\cline{3-4}
\colhead{} & 
\colhead{$\nu$} & 
\colhead{HPBW} & 
\colhead{PA} &
\colhead{$\Delta\nu$} &
\colhead{Resolution} &
\colhead{Noise}
\\
\colhead{Observation} & 
\colhead{(GHz)} & 
\colhead{(arcsec)}&
\colhead{(deg)}& 
\colhead{(MHz)} & 
\colhead{(km~s$^{-1}$)} &
\colhead{$\!\!\!\!$(mJy~beam$^{-1}$)} 
}
\startdata
Continuum 
 &  222.31   & $3.3\times1.5$ & 79 & 4000 & & 1.8 \\
\form\ \JK{3}{0,3}{2}{0,2} 
  & 218.2222 & $3.1\times1.8$ & 76 & 104 & 0.28 & 175 \\
DCN \J{3}{2} 
 & 217.2385  & $3.2\times2.1$ &76& 104 & 1.07 & 78 \\
SiO \J{5}{4} 
 & 217.1050  & $3.2\times2.1$ & 76 & 104 & 1.12 & 60 \\
CN \J{2}{1} \J{5/2}{3/2}\ \J{7/2}{5/2}\tablenotemark{a}
 & 226.8748  & $3.1\times2.0$ & 76 & 104 & 1.07 & 65 \\
CN \J{2}{1} \J{3/2}{1/2}\ \J{5/2}{3/2}\tablenotemark{a}
 & 226.6595  & $3.1\times2.0$ & 76 & 104 & 1.07 & 88 \\
\enddata
\tablenotetext{a}{CN N,J,F hyperfine transitions }
\end{deluxetable}

\begin{deluxetable}{lllccll}
\tablecaption{Dust properties at 1.35~mm\tablenotemark{a}\label{tdust}}
\tablewidth{0cm}
\tablehead{
\colhead{} & 
\multicolumn{2}{c}{Positions} & 
\colhead{$I_{\nu}$}& 
\colhead{$S_{\nu}$} &
\colhead{Deconvolved Size} &
\colhead{PA} 
\\
\cline{2-3}
\colhead{Source} & 
\colhead{$\alpha$(J2000)} & 
\colhead{$\delta$(J2000)} & 
\colhead{(mJy~beam$^{-1}$)}& 
\colhead{(mJy)} &
\colhead{(arcsec)} &
\colhead{(deg)} 
}
\startdata
SMA 1 
& $19^{\rm h}17^{\rm m}53\fs884$ 
& $19\arcdeg12'18\farcs51$ 
& 29.9$\pm$1.8 &45$\pm$4 
& $(2.0\pm0.3)\times (\la0.6)$
& $56\pm9$ \\
SMA 2 
& $19^{\rm h}17^{\rm m}53\fs694$ 
& $19\arcdeg12'19\farcs68$ 
& 33.7$\pm$1.8 & 59$\pm$4 
& $(2.3\pm0.2)\times(1.0\pm0.2)$ 
& $73\pm6$ \\
\enddata
\tablenotetext{a}{ Values corrected by the primary beam of the SMA antennas.}
\end{deluxetable}


\begin{table}
\caption[]{Masses of the different components in L723}
\label{tmass} 
\[
\begin{tabular}{lccc}
\hline
           & $S_{\nu}({\rm 1.35mm})$ & $T_{\rm dust}$ & $M$ \\
Source & (mJy)  & (K) & (M$_{\odot}$) \\
\hline
All$^a$& 128$\pm6$ & 25 & 0.24 \\
SMA~1  &  45$\pm4$ & 25 & 0.09 \\
SMA~2  &  59$\pm4$ & 25 & 0.11 \\
WHS    &  $\le5.7$ & 25 & $\le0.011$ \\
\hline
\end{tabular}
\]
    \begin{list}{}{}
\item[$^a$] To estimate the total mass we took into account the contribution of 
the mass from the 24~mJy excess of emission of the Natural map (obtained 
adopting a temperature of 20~K).
     \end{list}
\end{table}

\begin{table}
\caption{\form\ \JK{3}{0,3}{2}{0,2}}\label{tform}
\centering
\begin{tabular}{lccccc}
\hline
      &  $I_{\nu}$  &  $\int I_{\nu} dv$   & $v_{\rm LSR}$  & $\Delta v$ \\
Source&    (K)      &   (K\kms)    & (km~s$^{-1}$)  & (km~s$^{-1}$) \\
\hline
SMA 1 & 4.0$\pm$0.7 &  8.7$\pm$1.0 & 11.47$\pm$0.11 & 2.0$\pm$0.3\\
SMA 2 & 8.7$\pm$0.7 & 14.6$\pm$0.8 & 11.64$\pm$0.04 & 1.6$\pm$0.1\\
\hline
\end{tabular}
\end{table}

\begin{table}
\caption{Best \form\ Models}\label{tRatran}
\centering
\begin{tabular}{lccc}
\hline
&& $R_{in}$ & $R_{out}$ \\
Source & $X[$p--\form$]$ & (AU) &  (AU) \\
\hline
SMA 1 & $(8$--$30) \times10^{-11}$ & 30--120 & 600--5000 \\
SMA 2 & $(3$--$10) \times10^{-10}$ & 30--240 & 300--600 \\
\hline
\end{tabular}
\end{table}

\end{document}